\documentclass[twocolumn,showpacs,preprintnumbers,prb]{revtex4-1}
\usepackage{dcolumn}
\usepackage{amsmath}
\usepackage{amsfonts}
\usepackage{amssymb}
\usepackage{calligra}
\usepackage{bm}
\usepackage{graphicx}
\usepackage{float}
\usepackage{subfigure}
\usepackage{stackengine}

\newcommand{\e}[1]{e^{#1}} 
\newcommand{\avg}[1]{\left< #1 \right>} 
\newcommand{\abs}[1]{| #1 |} 
\newcommand{\ket}[1]{| #1 \rangle} 
\newcommand{\bra}[1]{\langle #1 |} 

\DeclareMathAlphabet{\mathcalligra}{T1}{calligra}{m}{n}

\begin{document}

\title{Single-particle Entanglement of the Entanglement Hamiltonian
  Eigen-modes}

\author{Mohammad Pouranvari}

\affiliation{School of Particles and Accelerators Institute research
  in fundamental science (IPM). P.O. Box 19395-5531, Tehran, Iran}

\date{\today}

\begin{abstract}
  Single-particle entanglement entropy (SPEE) is calculated for
  entanglement Hamiltonian eigen-mode in a one-dimensional free
  fermion model that undergoes a delocalized-localized phase
  transition. In this numerical study, we show that SPEE of
  entanglement Hamiltonian eigen-mode has the same behavior as EPEE of
  Hamiltonian eigen-mode at the Fermi level: as we go from delocalized
  phase toward localized phase, SPEE of both modes decreases in the
  same manner. Furthermore, fluctuations of SPEE of entanglement
  Hamiltonian eigen-mode -- which can be obtained through the
  calculation of moments of SPEE -- signature very sharply the phase
  transition point. These two modes are also compared by calculation
  of single particle R\'{e}yni entropy (SPRE). We show that SPEE and
  SPRE of entanglement Hamiltonian eigen-mode can be used as a phase
  detection parameter.
\end{abstract}

\maketitle

\section{Introduction}\label{Introduction}
Entanglement as a pure quantum concept with no classical counterpart,
has been used as a phase detection parameter\cite{ref:Horodecki}. It
is borrowed from quantum information science, and people in condensed
matter physics found it useful to distinguish different
phases\cite{ref:chuang, ref:Ekert, ref:Steane, ref:Gisin,
  ref:osterloh}. I.e. its behavior depends on the phase of the
system. Specially, for a delocalized-localized phase transition, the
concept of entanglement entropy (EE) is useful.\cite{ref1, ref2, ref3,
  ref4, ref:pouryuhui} In the delocalized phase, where the system is
extended over many sites, we expect large correlation in the system,
and thus EE -- which indirectly measures the correlation among the
system -- is larger than when the system is localized. R\'{e}yni
entropy (RE) is another measure of entanglement in a system, by which
people distinguish localized from delocalized phases.\cite{ref11,
  ref12, ref13} However, beside the entanglement, there are other
source of information contained in the reduced density
matrix. Eigen-values of the entanglement Hamiltonian are also another
way to distinguish different phase.\cite{ref:haldane} In addition,
eigen-modes of the entanglement Hamiltonian also carry physical
information.\cite{ref:pouranvariyang2,ref:pourafshin}

Let's review the concept of entanglement. For a free fermion
Hamiltonian, we obtain single-particle eigen-modes of the Hamiltonian,
and the ground state of the system $\ket{\psi}$ will be the Slater
determinant of filled single-particle eigen-mode up to Fermi level.
We know that all physical information contained in the state can also
be understood using the density matrix which is defined as
$\rho = \ket{\psi} \bra{\psi}$. Now, we consider a system divided into
two subsystems. For each subsystem, we can obtain a \emph{reduced}
density matrix, which is obtained by tracing over the other
subsystem. In this paper we consider a lattice system with $N$ sites
and divide the system into two equal parts: subsystem $A$ is from site
number $1$ to site number $N_A$ and the rest is subsystem $B$. Thus,
the reduced density matrix of subsystem $A$ for example, is obtained
by tracing over sites of subsystem $B$: $\rho_A = tr_B \rho$. Then,
entanglement entropy is $EE=-tr [\rho_A \log \rho_A]$. Note that
entanglement entropy is defined for a many-body state of the
system. In a free fermion lattice system, which we focus on in this
study, we can write the reduced density matrix as $\e{-H_{ent}}$ in
which the $H_{ent}$ is a free fermion Hamiltonian and called
\emph{entanglement Hamiltonian}. This procedure can be done to
calculate the reduced density matrix for subsystem $B$ as well.

In this paper, the single-particle eigen-modes of the entanglement
Hamiltonian are considered. Note that for each entanglement
Hamiltonian eigen-mode of subsystem $A$, there is a counterpart
eigen-mode in subsystem $B$. To obtain a mode that characterizes the
\textbf{entire} system, we attach these two
eigen-modes. Ref. [\onlinecite{ref:klich}] introduces a method for
attaching these two modes together. For a system with size $N$ that is
divided into two equal subsystems, we have $N/2$ of such single
particle eigen-modes for each subsystem. But, one of them is
particularly very important. Note that each eigen-mode of the
entanglement Hamiltonian, correspond to an eigen-value, that is the
corresponding entanglement energy. One of these eigen-values that is
closest to zero has the largest contribution to the EE. The
corresponding eigen-modes in two subsystems are attached together to
make a mode that is called maximally entangled mode (MEM) for the
whole system.

In recent studies, we showed that MEM has physical information very
similar to those information we can obtain from eigen-mode of the
Hamiltonian at the Fermi level, $\ket{E_F}$. It is shown that both MEM
and $\ket{E_F}$ are extended in the delocalized phase and both are
localized in localized phase and thus, by studying the MEM behavior we
can distinguish different phases: MEM introduces another way of
studying the behavior of the system.\cite{ref:pouranvariyang1,
  ref:pouranvariyang2, ref:pouranvariyang5}

For a single-particle eigen-mode, it is possible to define
single-particle entanglement entropy (SPEE)(see below for
definition). And since the MEM is a single-particle eigen-mode, we can
obtain its SPEE, as we do for the single-particle eigen-mode of the
Hamiltonian at the Fermi level. On the other hand, we have seen that
the behavior of the MEM is similar to the behavior of Hamiltonian
eigen-mode at the Fermi level. We conjecture that there is a physics
in the MEM that can be captured by measuring its correlation, through
the calculation of its entanglement. Although the MEM and $\ket{E_F}$
are not the same, but since the behavior of both are similar, we
expect that the correlation in the MEM to have the same trend as the
correlation of the $\ket{E_F}$. As a single-particle eigen-mode, we
can calculate the single-particle entanglement entropy of MEM to
obtain the correlation information in the MEM. In this paper, we show
that SPEE of the MEM, distinguishes different phases and it locates
the phase transition point, and thus can be used as a phase detection
parameter.

Summary of our results are the followings: SPEE of MEM has the same
behavior as SPEE of $\ket{E_F}$. As we go from delocalized phase to
localized phase, SPEE of both modes decreases. In addition, amount of
fluctuations in SPEE of MEM can be used as a signature of the phase
transition point: in the delocalized phase moments of the SPEE of MEM
is very small, but its magnitude sharply increases at the phase
transition point. Furthermore, we calculate the single-particle RE of
both these modes and show that they have the same behavior, i.e. both
modes have same entanglement information about the system.

Paper is structured as follows: first, in section \ref{s1}, we explain
the concept of single-particle EE along with the models we employ in
this paper, and then we compare the SPEE of $\ket{E_F}$ and MEM. In
section \ref{s2}, we use the notation of RE as another comparison of
these two modes. Paper is finished with a conclusion in section
\ref{s3}.

\section{Single-particle Entanglement entropy of MEM}\label{s1}
In this paper, single-particle eigen-modes of entanglement Hamiltonian
are considered and they should not be confused with the many-body
states of the system of the original Hamiltonian. In what follows, we
explain the SPEE which will be applied to single-particle MEM as well
as to single-particle $\ket{E_F}$. As explained in
Ref. [\onlinecite{ref:jia, ref:zanardi}] to define the SPEE, we use
occupation number basis. For a lattice system with size $N$:
\begin{equation}
  \ket{\psi} = \sum_{i=1}^{N} \psi_i \ket{1}_i \bigotimes_{j\ne i} \ket{0}_j,
\end{equation}
where $\ket{\psi}$, can be the $\ket{E_F}$ or the MEM. We divide the system
into two equal parts, $A$ and $B$. We can define:
\begin{eqnarray}
  \ket{1}_A &=& \sum_{i=1}^{N_A} \psi_i \ket{1}_i \bigotimes_{j\ne i}\ket{0}_j, \\
  \ket{0}_A &=& \bigotimes_{i=1}^{N_A}\ket{0}_i.
\end{eqnarray}

To obtain the reduced density matrix for subsystem $A$, we trace over
sites in $B$ and we obtain:
\begin{equation}
  \rho_A = \ket{1}_A\bra{1} + p_B \ket{0}_A\bra{0}, 
\end{equation}
where:
\begin{eqnarray}\label{pa}
  p_A  &=&  \sum_{i=1}^{N_A} \abs{\psi_i}^2,\\
  p_B  &=&  \sum_{i=N_A+1}^{N}  \abs{\psi_i}^2 = 1-p_A, \label{pb}
\end{eqnarray}
and finally we obtain the SPEE:
\begin{equation}\label{EE}
  EE = -tr \rho_A \log\rho_A = -(p_A \log p_A + p_B \log p_B).
\end{equation}

We note that the above mentioned procedure to calculate the
single-particle entanglement entropy can be applied to any
single-particle wave-function in lattice system; it can be applied to
$\ket{E_F}$ as well as to MEM.

To verify our idea, we employ power-law random banded matrix model
(PRBM)\cite{ref:mirlinprbm} that is a one-dimensional long range
hopping model with the following Hamiltonian:
\begin{equation}\label{Hhij}
  H = \sum_{i,j=1}^N h_{ij}c^{\dagger}_i c_j
\end{equation}
in which $N$ is the system size and $c^{\dagger}_j(c_j)$ is the
creation (annihilation) operator for site $j$ in the second
quantization. Matrix elements $h_{ij}$ are random numbers that are
distributed by a Gaussian distribution. The mean value of the
distribution is zero and it has the following variance:
\begin{equation} \label{Hobc}
  \avg {\abs{h_{ij}}^2} = \left[1+(\frac{|i-j|}{b})^{2a}\right]^{-1}
\end{equation}
Where $b$ is a parameter, by which we can tune hopping amplitudes. In
the regime of $b \ll 1$ we approach to the nearest-neighbor case; on
the other hand, when $b \gg 1$, all hopping amplitudes are
non-zero. In this paper we set $b=1$.\cite{ref:mirlinfootnote} To
avoid the effect of finite size of system, we choose periodic boundary
condition in the Hamiltonian of Eq. (\ref{Hobc}), where we replace
$i-j$ with the chord length and thus the Hamiltonian becomes:
\begin{equation} 
  \avg {\abs{h_{ij}}^2} = \left[{1+\left(\frac{\sin{\pi
            (i-j)/N}}{b \pi /N}\right)^{2a}}\right]^{-1},
\end{equation}
as it is proved numerically and analytically in
Ref. [\onlinecite{ref:mirlinprbm}], this system is localized for $a>1$
and it is extended in the regime of $a<1$. There is a phase transition
at the point $a=1$ regardless the value of $b$. This model is
important since there is a parameter $b$ in this model that can be
tuned in a way that it resembles other typical
models\cite{ref:jose,ref:balatsky,ref:altshuler,ref:ponomarev,ref:casati,ref:mirlin2000},
and specially it can be tuned to have similar behavior like the three
dimensional Anderson model. This model has attracted much attention
and has been used in several recent studies (see for example
Ref. [\onlinecite{refP1, refP2, refP3, refP4}]). Because of such
features, we choose this model to verify our ideas.

\begin{figure}
  \centering
  \includegraphics[width=0.5\textwidth]{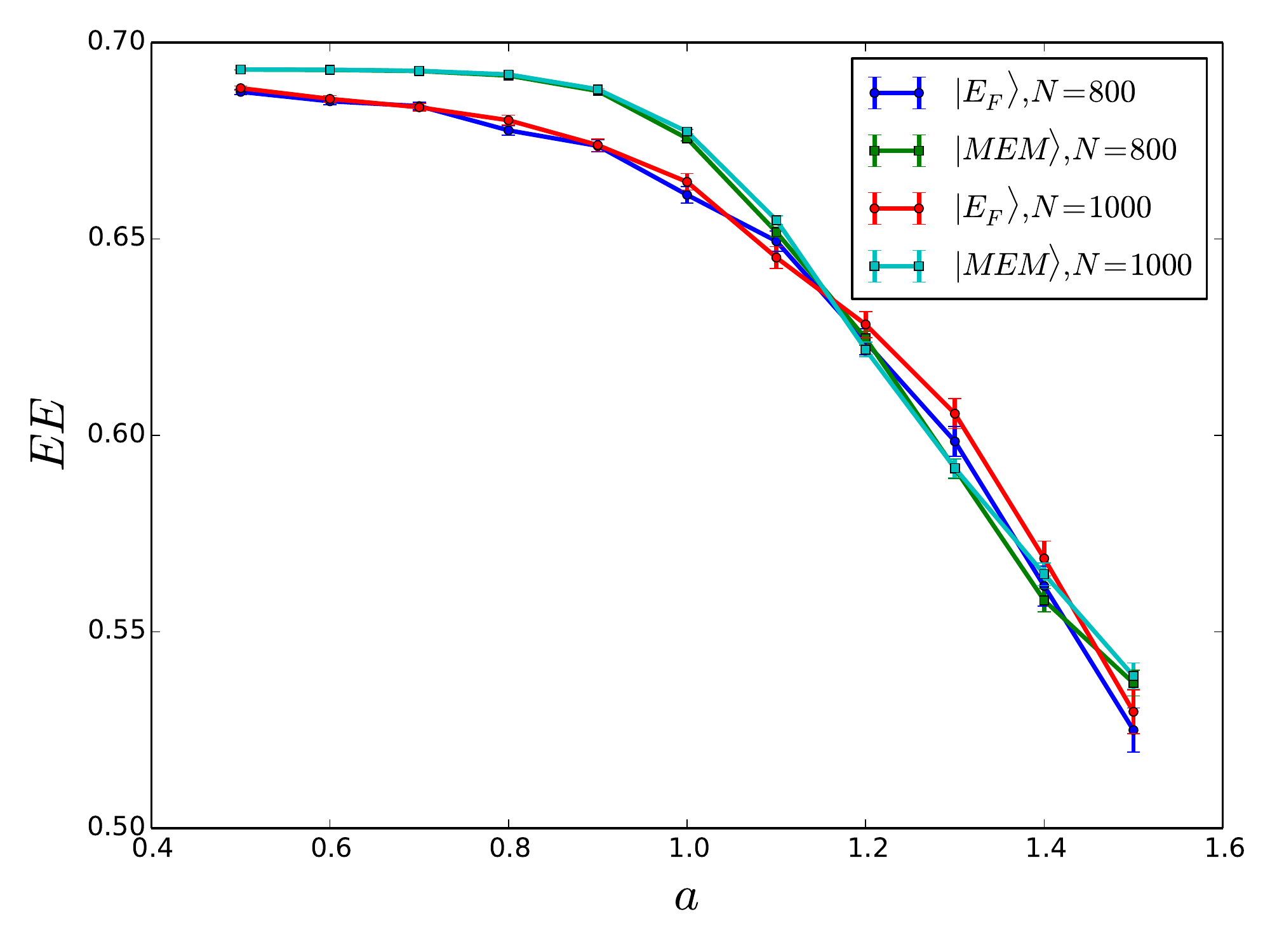}
  \caption{Single particle entanglement entropy of  $\ket{E_F}$ and
    $\ket{MEM}$ as a function of $a$. It decreases for both of them as
    we go to the localized phase. For each point we have $2000$
    samples. Standard error are plotted in figure by a vertical line.
    \label{fig:PRBM_EE_a}}
\end{figure}

We calculate the SPEE of the MEM and $\ket{E_F}$ by
Eq. (\ref{EE}). Results are presented in
Fig. \ref{fig:PRBM_EE_a}. Although the SPEE of both are not exactly
the same, both have the same trend. In the delocalized phase, SPEE is
almost constant and as we approach to the the localized phase, SPEE
decreases. This result makes sens, since as we approaches to the
localized phase, the amount of correlation and thus SPEE decreases.

\begin{figure}
  \centering
  \includegraphics[width=0.5\textwidth]{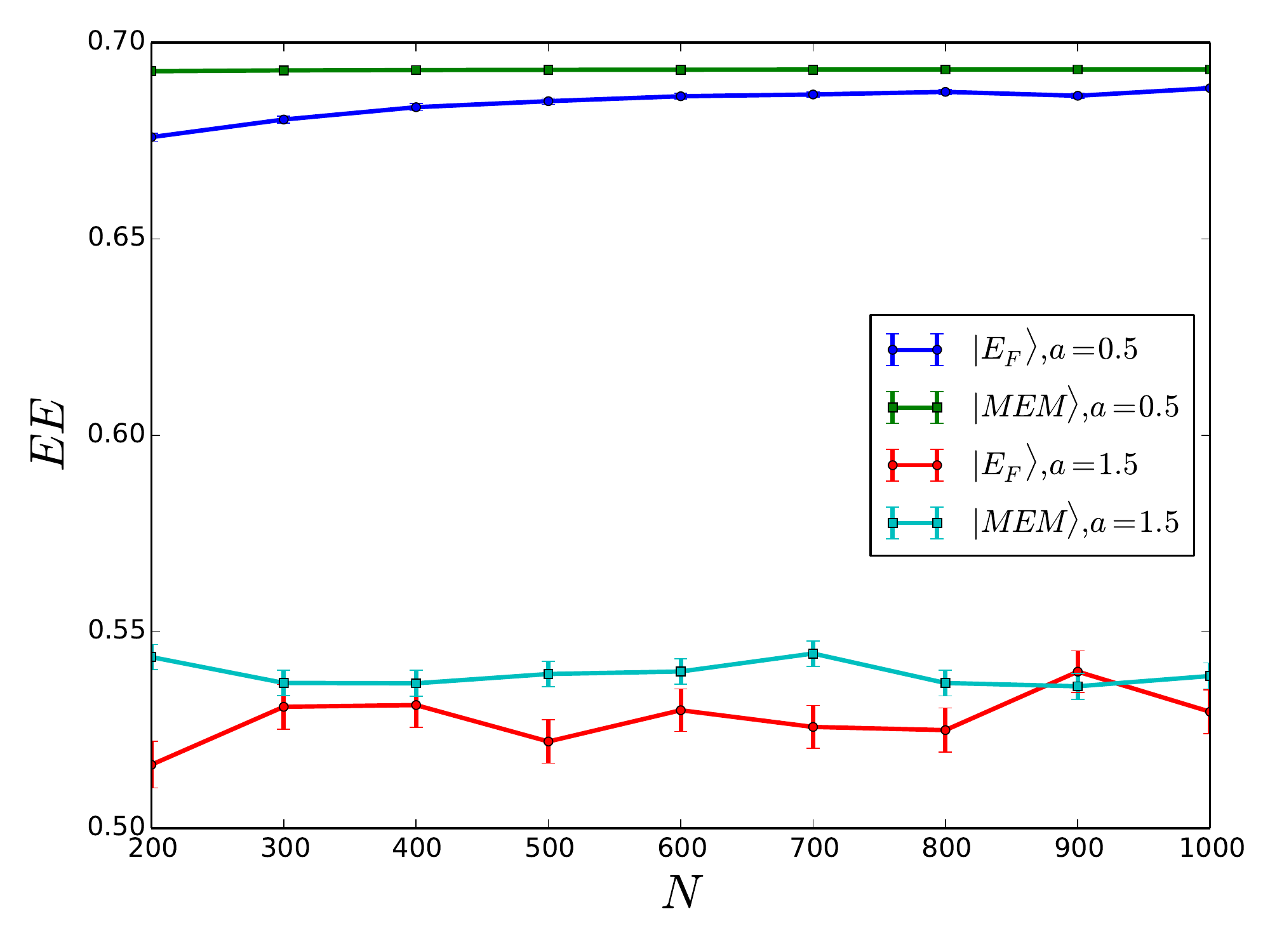}
  \caption{Single particle entanglement entropy of $\ket{E_F}$ and
    $\ket{MEM}$ as a function of system size $N$. In the delocalized
    phase is around $\ln{(2)}$. }
  \label{fig:PRBM_EE_N}
\end{figure}

On the other hand, we plot SPEE of both $\ket{E_F}$ and MEM as system
size $N$ changes (see Fig. \ref{fig:PRBM_EE_N}). SPEE for MEM and
$\ket{E_F}$ are approximately constant as $N$ changes, thus we
do not need a very large system size to verify the phase of the
system.

\begin{figure}
  \centering
  \begin{subfigure}{}%
    \includegraphics[width=0.22\textwidth]{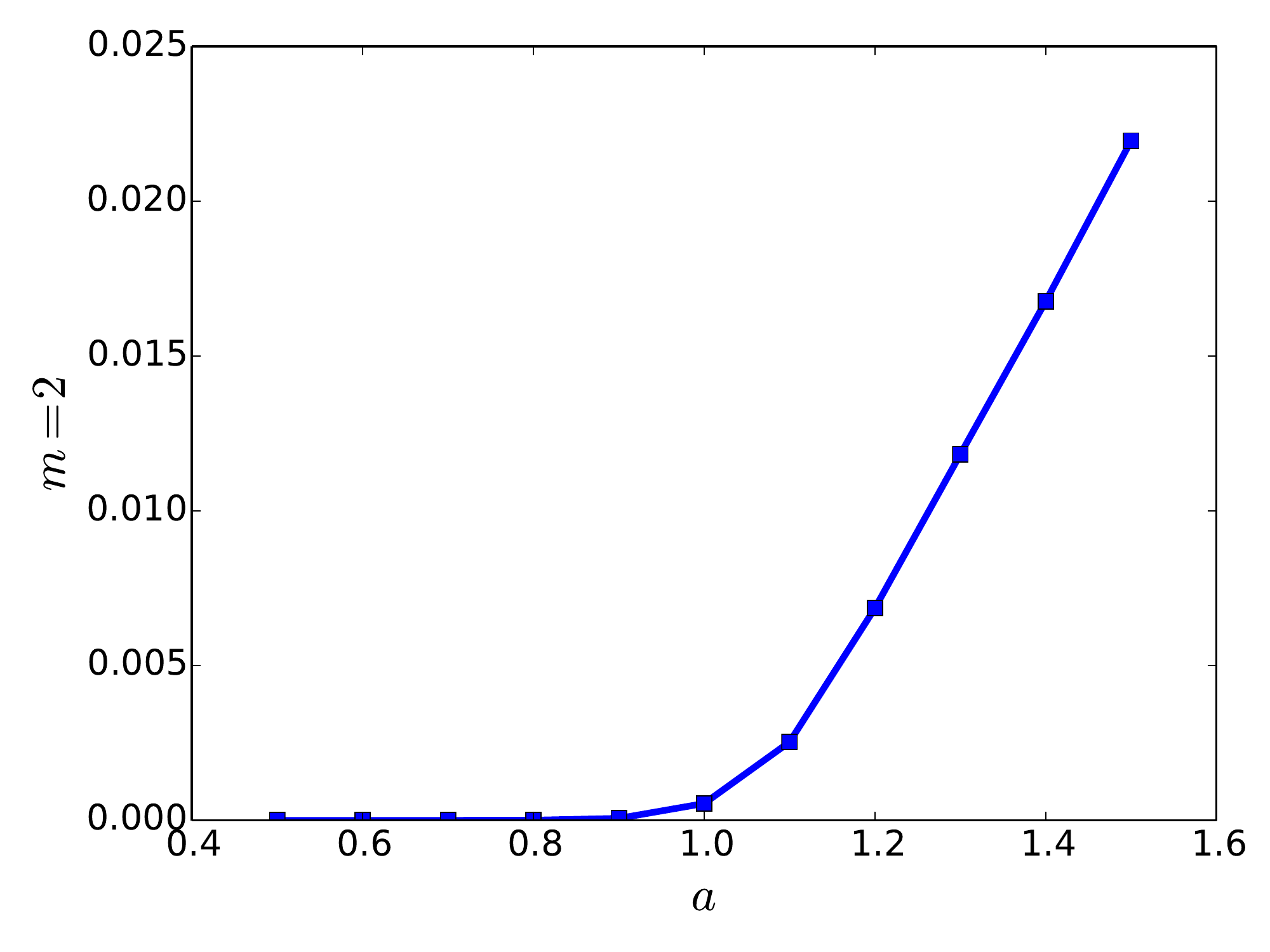}
  \end{subfigure}%
  ~%
  \begin{subfigure}{}%
    \includegraphics[width=0.22\textwidth]{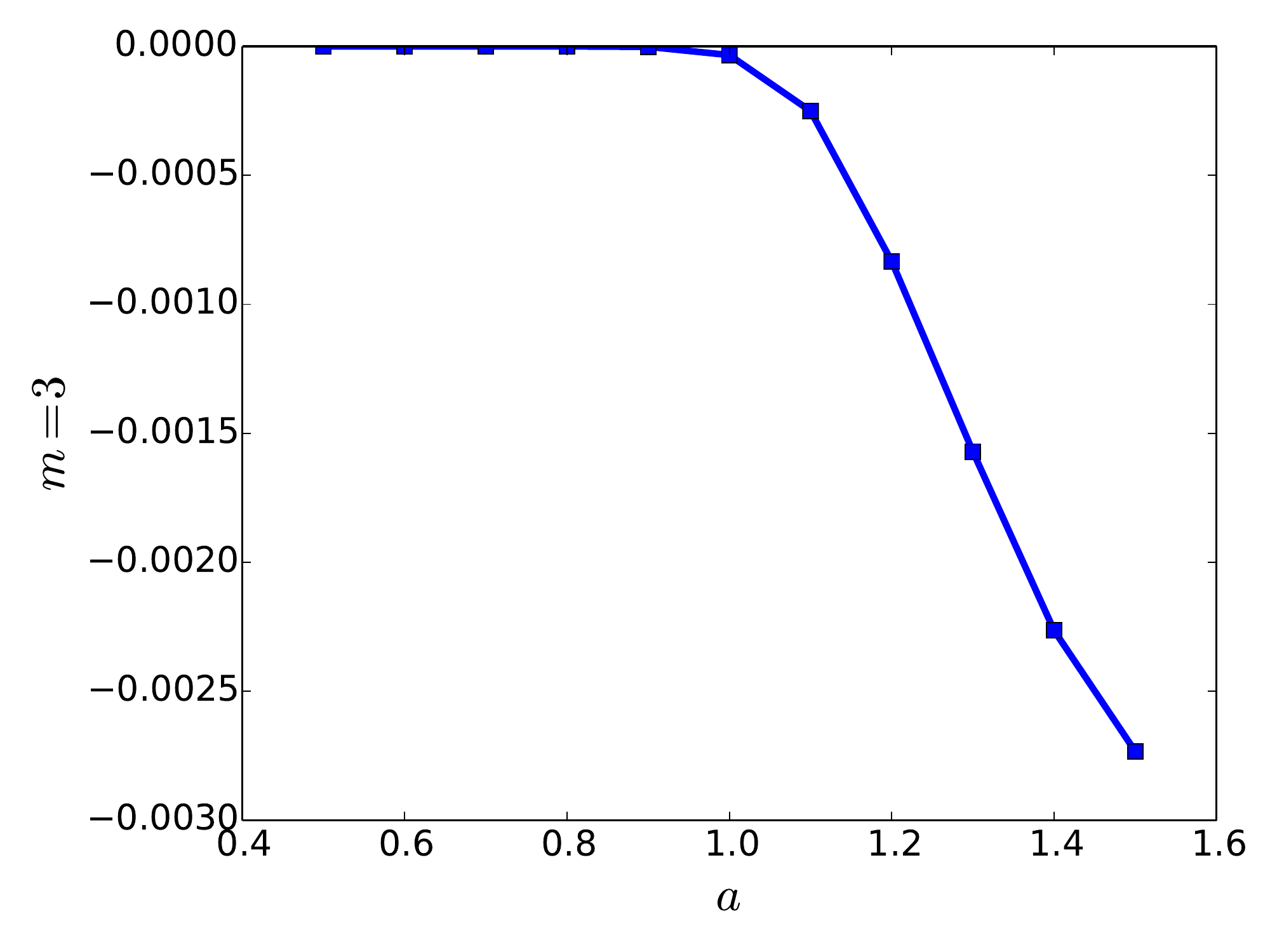}
  \end{subfigure}
  
  \begin{subfigure}{}%
    \includegraphics[width=0.22\textwidth]{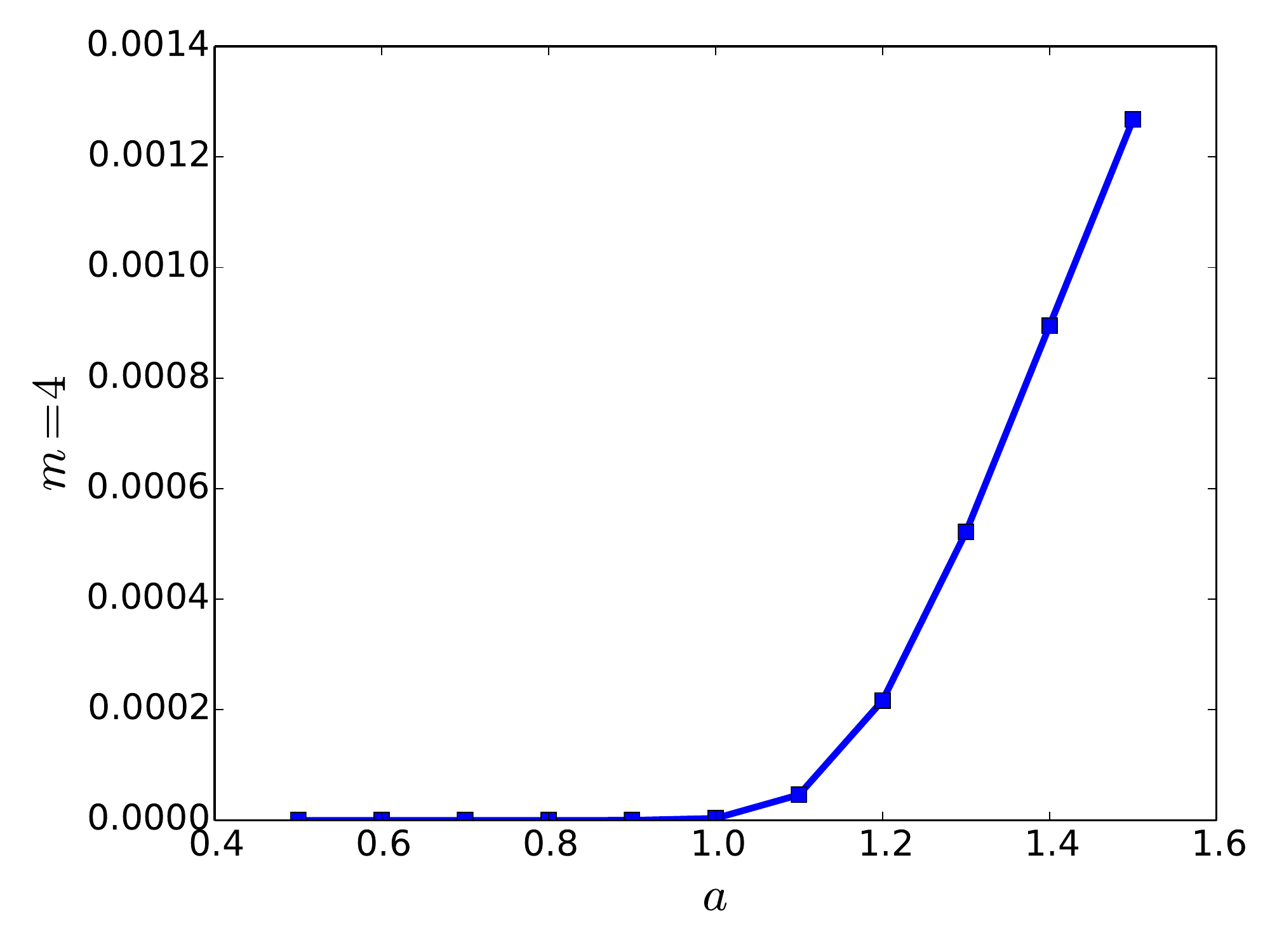}
  \end{subfigure}
  ~%
  \begin{subfigure}{}%
    \includegraphics[width=0.22\textwidth]{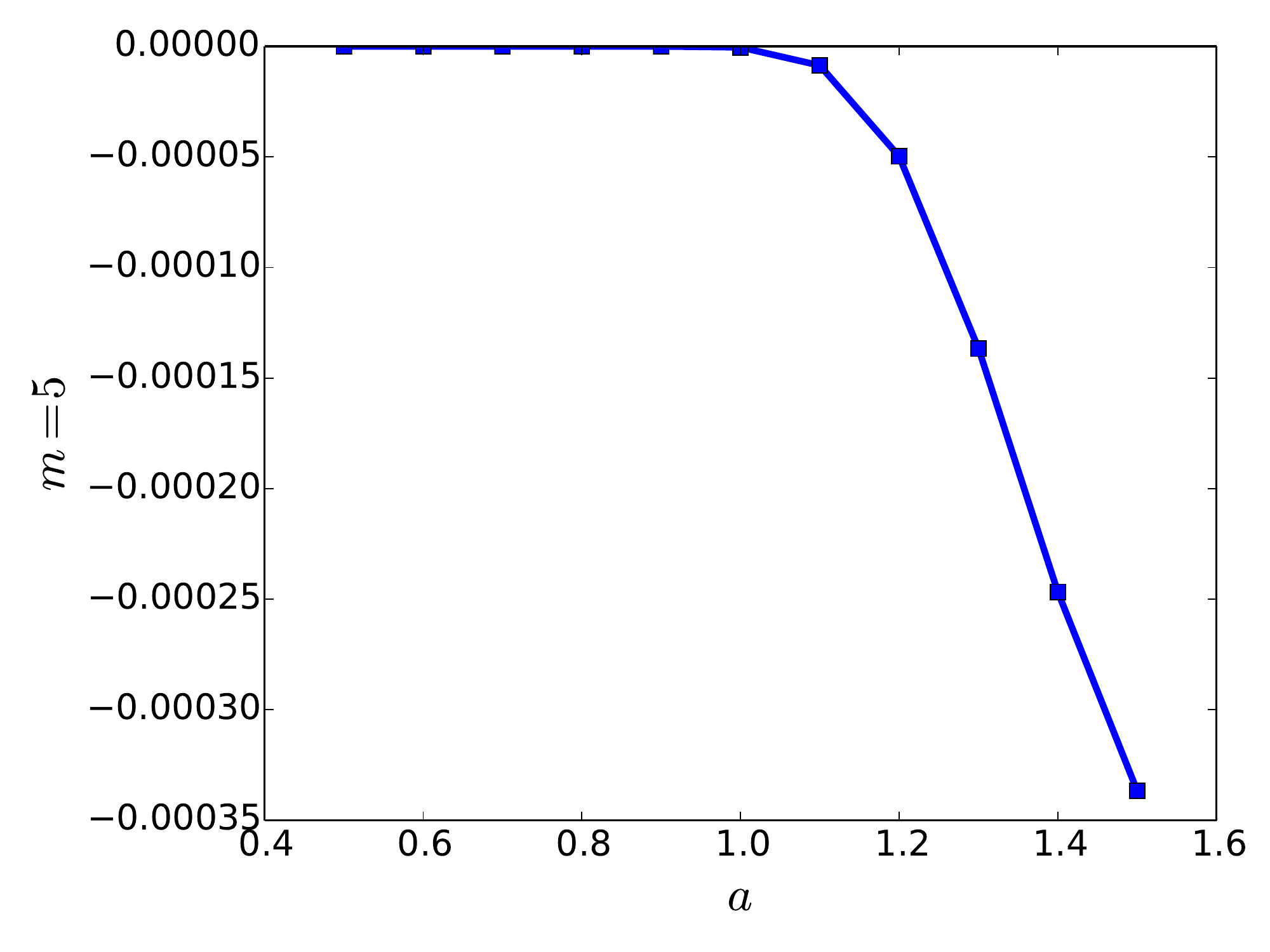}
  \end{subfigure}
  \caption{Moments of single-particle entanglement entropy for MEM as
    we change $a$. Moments for $m=2,3,4,5$ clearly distinguish
    different phase. System size is fixed and equal to $N=1000$. These
    plots show that fluctuations of SPEE of MEM signature the phase
    transition.
    \label{fig:momentEE}}
\end{figure}

PRBM model undergoes a disorder phase transition, i.e. by increasing
disorder in system, it goes from delocalized phase to localized phase
where quantum fluctuations are dominant. These fluctuations should be
seen in observable quantities in system. To see how fluctuations are
seen in the SPEE, we use moments of SPEE. Moments, $m_k$, of random
numbers $\{x\}$ with mean value $<x>$ are defined as below:
\begin{equation}
  m_k = \frac{1}{n_s} \sum_{i=1}^{n_s} (x_i-<x>)^k
\end{equation}
where $n_s$ is the number of random numbers. Different moments of SPEE
of MEM are plotted in Fig. \ref{fig:momentEE} for $m=2,3,4,5$. As we
can see, different moments that are a measure of fluctuations of SPEE
in system, show very sharply the phase transition point: Moments are
zero in the delocalized phase and they are non-zero in the localized
phase. Thus, fluctuations of the SPEE of MEM can distinguish different
phase.

\section{Single Particle R\'{e}yni Entropy of MEM} \label{s2}

\begin{figure}
  \centering
  \includegraphics[width=0.5\textwidth]{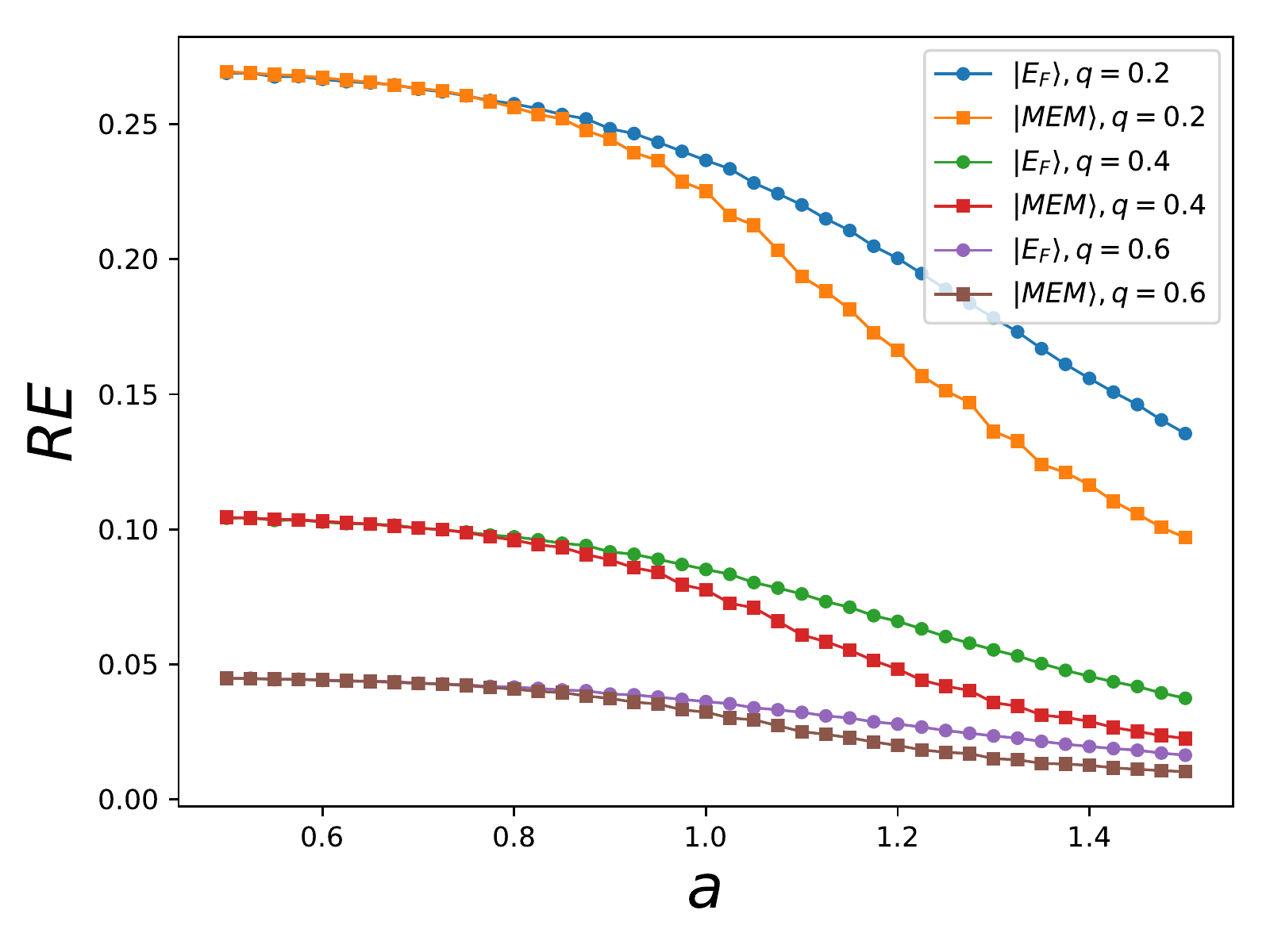}
  \caption{Single particle R\'{e}yni entropy of $\ket{E_F}$ and
    $\ket{MEM}$ as a function of $a$. We see that both have same
    trend. System size is fixed to $N=500$ and for each point we have
    $1000$ samples.}
  \label{fig:RE_Q_PRBM}
\end{figure}

Beside the single-particle entanglement entropy that contains
information about correlation of the system and is a tool to
distinguish different phases, we can also use the other related
measure, namely the R\'{e}yni entropy. Similar to EE, RE can also be
calculated for a many-body state as well as for a single-particle
state. Here we apply single-particle R\'{e}yni entropy (SPRE) to the
single-particle MEM and $\ket{E_F}$. The advantage of RE over EE is
that, we can obtain more complete physical information by calculating
different orders of R\'{e}yni entropy. Here, we calculate the RE of
MEM and compare it with RE of $\ket{E_F}$ to show that, both have
similar entanglement properties from the perspective of other
measurement of entanglement, namely the RE.

because we have the freedom of choosing size of the subsystem, we
choose subsystem $A$ to be a single site and the rest of the system as
subsystem $B$. Since no site has privilege over other sites, we
average over sites of the system. SPRE then will be\cite{ref:chen}:
\begin{eqnarray}
  RE_{q} & = &\frac{1}{N} \frac{1}{1-q} \log{ tr \rho_A^q}\\
         &=& \frac{1}{N} \frac{1}{1-q} \log{(p_A^q + p_B^q)}
\end{eqnarray}
where $p_A$ and $p_B$ are defined in Eqs. (\ref{pa}) and
(\ref{pb}). In Fig. \ref{fig:RE_Q_PRBM}, we plot SPRE of $\ket{E_F}$
and MEM for a fixed system size as we change $a$. As we can see from
this figure, SPRE for both eigen-mode has similar behavior: SPRE is
decreasing as we go from delocalized phase to localized phase. Thus,
in the light of RE, both MEM and $\ket{E_F}$ have similar entanglement
properties in the delocalized-localized phase transition.

\section{conclusion}\label{s3}
In this paper, we introduced a novel phase transition
characterization, namely the single-particle entanglement and
R\'{e}yni entropy of the MEM of Hamiltonian eigen-mode.  By using a
free fermion lattice model that undergoes delocalized-localized phase
transition, we verified numerically that SPEE of MEM clearly
distinguish different phases . In the delocalized phase, SPEE is very
close to $\ln 2$ and it decrease in localized phase. Fluctuations of
the SPEE of MEM was another phase detection characterization we
introduced in this paper. In addition, to show that both MEM and
$\ket{E_F}$ have similar entanglement properties, we used another
measure of entanglement, namely the RE. We showed that SPRE for MEM
and $\ket{E_F}$ have similar behavior in delocalized and localized
phase.

\section{Acknowledgments}
This research is supported by National Merit Foundation of Iran, and
Institute for Research in Fundamental Sciences (IPM).

\end{document}